\documentclass[prl,twocolumn]{revtex4}
\usepackage{graphicx}
\usepackage{amsmath}
\usepackage{amssymb}
\usepackage{bm}
\usepackage{mathrsfs}
\usepackage{color}
\usepackage{slashed}
\usepackage{dcolumn}
\usepackage{float}
\usepackage{natbib}

\newcommand{\eqf}[1]{\begin{equation}\begin{split}#1\end{split}\end{equation}}

\begin{document}

\title{The $Z^0$-tagged jet event asymmetry in heavy-ion collisions \\ 
at the CERN Large Hadron Collider}

\author{R. B. Neufeld}
%\email{neufeld@lanl.gov}
\address{Los Alamos National Laboratory, Theoretical Division, MS B238, Los Alamos, NM 87545, U.S.A.}

\author{I. Vitev}
%\email{ivitev@lanl.gov}
\address{Los Alamos National Laboratory, Theoretical Division, MS B238, Los Alamos, NM 87545, U.S.A.}

\date{\today}

\begin{abstract}
Tagged jet measurements provide a promising experimental channel to quantify the 
similarities and differences in the mechanisms of jet production in proton-proton
and nucleus-nucleus collisions. We present the first calculation of the transverse
momentum asymmetry of $Z^0/\gamma^*$-tagged jet events in $\sqrt{s}=2.76$~TeV reactions 
at the LHC. Our results combine the ${\cal O}(G_F\alpha_s^2)$ perturbative cross 
sections with the radiative and collisional processes that modify parton showers 
in the presence of dense QCD matter. We find that a strong asymmetry is generated 
in central lead-lead reactions that has little sensitivity to the fluctuations of 
the underlying soft hadronic background. We present theoretical model predictions
for its shape and magnitude. 
\end{abstract}

\maketitle

Jet production in energetic particle collisions
is one of the most powerful 
channels through which to test and advance perturbative Quantum Chromodynamics
(QCD)~\cite{Sterman:1977wj}. This is particularly true at the CERN Large Hadron 
Collider (LHC), where  the available large 
center-of-mass energies  guarantee an abundant yield of high transverse momentum 
hadrons and jets~\cite{Olness:2009qd}. In relativistic heavy-ion collisions at the LHC, 
parton shower formation  and evolution are  modified by the hot and dense deconfined matter, 
or quark-gluon plasma (QGP), created  during the early stages of the interaction~\cite{jetty}.   
Consequently, the related jet observables can help quantify the properties of the QGP and 
differentiate between competing paradigms of jet  production and modification in ultra-relativistic
nuclear collisions.

In light of the above motivation, it is not surprising that quantitative theoretical 
description and experimental measurement of jet observables in heavy-ion collisions have 
become an important priority, both at the Relativistic Heavy-Ion Collider (RHIC) and the LHC~\cite{jets}. Recently, the ATLAS
and CMS collaborations reported a significant enhancement in the 
transverse momentum imbalance of di-jets produced in central lead-lead  (Pb+Pb) collisions at the LHC 
relative to the ones produced in proton-proton  (p+p) collisions~\cite{Aad:2010bu}. The broader distribution of 
the asymmetry variable, denoted $A_J$  and defined as 
\eqf{\label{ass}
A_J = \frac{{p_T}_1 - {p_T}_2}{{p_T}_1 + {p_T}_2} \,,
}
where ${p_T}_1$ and ${p_T}_2$ are the transverse momenta of the leading and subleading jets,
is reflective of the jet in-medium modification. Attempts to explain the observed asymmetry 
based on Monte Carlo simulations with a Pythia generated p+p baseline have been 
presented~\cite{Qin:2010mn}. Fixed order ${\cal O}(\alpha_s^3)$ 
pQCD calculations that include nuclear matter effects describe well both single
and di-jet production in nucleus-nucleus (A+A) reactions at the LHC~\cite{He:2011pd}.  
The importance of potentially large background fluctuations in heavy-ion collisions on the 
$A_J$ asymmetry distribution has been argued for in Ref.~\cite{Cacciari:2011tm}. 

Jets tagged with electroweak bosons offer a complementary channel to di-jet measurements and 
have the potential to circumvent some of the problems inherent in multi-jet observables.  
Because the tagging boson does not interact strongly, this channel has been proposed 
as an experimental avenue through which to constrain the initial energy of the associated 
jet~\cite{Srivastava:2002kg}. In both p+p and A+A reactions, however,
next-to-leading accuracy is necessary for quantitatively and even qualitatively accurate
description of tagged jet production~\cite{Neufeld:2010fj}.
In what follows we provide first theoretical predictions to ${\cal O}(G_F\alpha_s^2)$ for 
the asymmetry and nuclear modification factor of $Z^0/\gamma^*$-tagged jets in 
heavy-ion collisions at LHC energies. Our calculation includes both the radiative 
and collisional processes important for parton shower modification in a nuclear medium.  
The heavy-ion program at the LHC has now enabled the first experimental measurements 
of $Z^0$ boson production~\cite{Chatrchyan:2011ua}. Although differential 
tagged-jet measurements in this channel require significantly more statistics to become quantitatively 
rigorous, there is  ongoing effort on this front. Our predictions are 
complementary to this experimental program and will provide timely 
guidance to the qualitative expectations for all electroweak boson-tagged jet measurements.

\begin{figure}
\centerline{
\includegraphics[width = 0.95\linewidth]{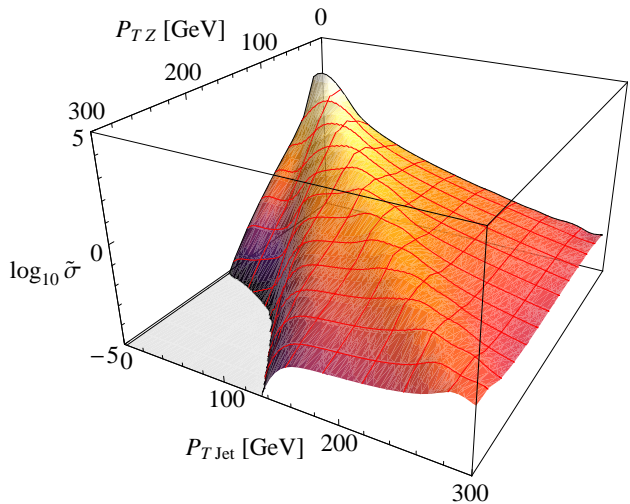}\hskip 0.03\linewidth
}
\caption{The ${\cal O}(G_F\alpha_s^2)$  cross section for $Z^0/\gamma^*$-tagged jets, 
where  $Z^0/\gamma^* \rightarrow \mu^+ \mu^- $ in proton-proton collisions at 
$\sqrt{s} = 2.76$ GeV ($R=0.4$).
}
\label{pp_lhc_tagged}
\end{figure}

In p+p reactions we evaluate the $Z^0/\gamma^*$-tagged jet production using 
the Monte Carlo for FeMtobarn processes (MCFM) code~\cite{Campbell:2002tg}.  
MCFM provides one-loop results for many QCD processes of interest to the hadron 
collider physics community. A principal channel used to measure $Z^0$s is their 
decay to di-leptons. All cross sections presented in this paper are for the 
$Z^0/\gamma* \rightarrow \mu^+\mu^-$. We implement the following acceptance cuts: 
$|y| < 2.5$ for both jets and final-state muons. We constrain the invariant mass of the 
muon pair in an interval around $M_z = 91.2$ GeV to fully contain the $Z^0$ peak. 
Jets are reconstructed using MCFM's built-in midpoint cone algorithm with a separation
parameter $R = \sqrt{\delta \phi^2 + \delta\eta^2}$. Standard 
$\mu_r = \mu_f  = \sqrt{M_Z^2+p_{T\,Z}^2}$ renormalization and factorization
scales and MSTW parton distribution functions are used~\cite{Martin:2009iq}.  
For presentation purposes we define the dimensionless double differential
cross section  
\begin{equation}
\tilde{\sigma} =  \left[\frac{\text{GeV}^2}{fb}\right] \, 
\frac{d\sigma^{\rm Z-jet}}{d p_{T\, Z} \, d p_{T\, \rm Jet}}
\label{brief}
\end{equation} 
and show $\log_{10} {\tilde \sigma}$ in Fig.~\ref{pp_lhc_tagged}  
for  $\sqrt{s}=2.76$~TeV  p+p collisions at the LHC. We have chosen 
a typical jet reconstruction parameter $R = 0.4$. The calculation was
performed in $\Delta p_{T\, Z} = 20$~GeV and  $\Delta p_{T\, \rm Jet} 
= 5$~GeV bins. The most important feature of
this cross section is how broad it is in the $(p_{T\, Z}, p_{T\, \rm Jet})$ 
plane~\cite{Neufeld:2010fj}, defined by the transverse momentum of the 
$Z^0$ boson and the transverse momentum of the jet. Its precise shape is 
determined by the parton level processes and the  $Z^0 \rightarrow \mu^+ \mu^- $ 
Dalitz decay kinematics.

%%%%%%%%%%%%%%%%%%%%%%%%%%%%%%%%%%%%%%%%%%%%%%%%%%%%%%%%%%%%%%%%%%%%%

In reactions with heavy nuclei, the inclusive and tagged jet production cross sections 
are modified by effects induced by the passage of the hard-scattering  partons and the
resulting parton showers through the strongly-interacting medium created in these reactions.   
Initial-state cold nuclear matter effects for $Z^0$ production at $\sqrt{s} = 2.76$~TeV were 
shown to be small both experimentally and theoretically~\cite{Chatrchyan:2011ua}. 
Furthermore, initial-state effects described in~\cite{lhcrad} do not affect the asymmetry of 
di-jet or tagged jet events~\cite{He:2011pd}.  

Final-state quark-gluon plasma effects include medium-induced parton splitting and the 
dissipation of the energy of the parton shower through collisional interactions in the 
strongly-interacting matter. Medium-induced parton splitting factorizes 
from the hard scattering cross section
and enters observables as a standard integral convolution~\cite{Ovanesyan:2011kn}. 
In the limit when the sub-leading parton carries on average a small fraction
$x  \ll 1$ of the parent parton's large lightcone momentum ($p^+ = p^z + p^0$ 
for a parton traveling in the $z$ direction), these processes have a 
transparent energy loss interpretation. The magnitude and angular distribution of radiative
energy losses are here described by the reaction operator formalism~\cite{Vitev:2007ve}.  
Specifically, we use the fully differential bremsstrahlung spectra 
for hard quarks and gluons, averaged  over the collision geometry in central 
Pb+Pb reactions at the LHC, that have been  previously employed to discuss jet and 
particle production in heavy-ion collisions~\cite{lhcrad,He:2011pd}.  
The probability density that the hard  scattered quark or gluon  will lose a fraction 
of their  lightcone  momentum  $ \epsilon = \sum_i x_i $ due to multiple gluon emission 
(the summation is over emitted gluons), 
or a medium-induced parton shower, is also evaluated and denoted
$P_{q,g}(\epsilon)$, respectively.  

The collisional energy losses are here motivated by the work done in 
Ref.~\cite{Neufeld:2011yh}. The energy transferred from the induced parton shower 
to the nuclear medium is evaluated in the hard thermal loop approximation to 
leading logarithmic accuracy. We are careful to keep track of the color correlations 
between the constituents within the shower and find that rate of energy loss
is suppressed at timescales $\delta t \sim 1/m_D \theta$ relative to the 
naive superposition of two independent partons. Here $m_D= g_{\text{med}} T$ is the Debye 
screening mass for a gluon-dominated plasma, $g_{\text{med}}$ is the strong coupling 
constant that sets the rate of parton energy loss, and $\theta$ is the parton splitting 
angle.  For large-angle radiation, which is characteristic of medium-induced 
showers~\cite{jetty,jets}, this effect is small. Our simulations suggest that
the shower generated by the propagation of a 75~GeV gluon through medium can transfer 
as much as 20~GeV of its energy to the medium.    

To relate the generation of medium-induced parton showers and the dissipation of 
part of their energy in the QGP to experimental observables, we need to implement 
the effects of the jet reconstruction kinematics. Let us define by $f(\omega_{\rm min},R)$ 
the fraction of the energy that is simply redistributed inside the jet of radius 
$R$~\cite{jetty}. Here, $\omega_{\rm min}$ is a parameter that simulates the effects 
of collisional energy loss discussed above~\cite{He:2011pd}. 
A hard parton contributing a fraction $\epsilon$ of its transverse momentum 
$p_T$ to a medium-induced shower will produce a jet of  
$p_{T\, \rm Jet}  = [\, 1 - (1-f(\omega_{\rm min},R))\epsilon \, ] p_T$. 
The resulting cross section per binary collision (of $\langle N_{\rm bin} \rangle$ total) 
reads~\cite{Neufeld:2010fj}  
\begin{eqnarray}
\label{2Dquench}
\frac{1}{  \langle N_{\rm bin} \rangle }
\frac{ d\sigma_{AA}   }{ d p_{T\, Z} d p_{T\, \rm Jet}} &=& 
\sum_{q,g}  \int_0^1 d \epsilon
\frac{P_{q,g}(\epsilon)}{  1 - (1-f(\omega_{\rm min},R)) \epsilon   } \nonumber \\
&& \hspace*{-.1cm}\times \frac{ d\sigma^{q,g} 
\left( \frac{ p_{T\, \rm Jet} }{ 1 - (1-f(\omega_{\rm min},R)) \epsilon } \right)  }
 {dp_{T\, Z} d p_T } \;.
\end{eqnarray}
The physical meaning of Eq.~(\ref{2Dquench}) is that the observed tagged jet cross section 
in nucleus-nucleus reactions is a probabilistic superposition of
cross sections for jets of higher initial transverse energy. This excess energy is then 
redistributed outside of the jet due to strong final-state interactions.

\begin{figure}
\centerline{
\includegraphics[width = 0.95\linewidth]{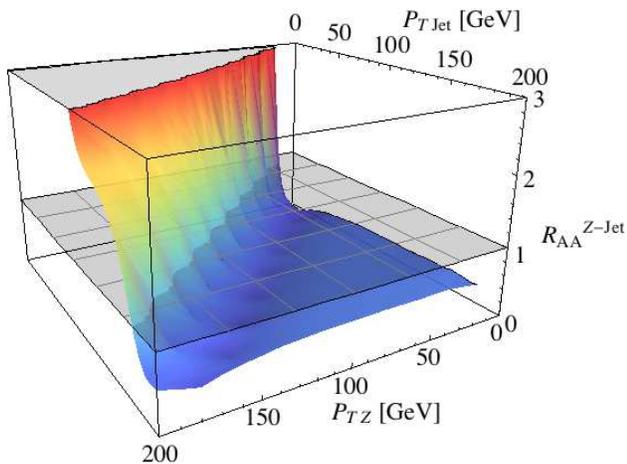}
}
\caption{The tagged jet nuclear modification factor, $R_{AA}^{\rm Z-jet}$, defined in  
Eq.~(\ref{RAAjet}), including both collisional and radiative energy loss effects. 
Our result is for $R=0.4$ and coupling between the jet and the medium given by 
$g_{\rm med} = 2$. 
}
\label{raa_lhc_tagged}
\end{figure}

We first consider the generalized jet nuclear modification factor, $R_{AA}^{\rm Z-jet}$, given by
\begin{eqnarray}
&&R_{AA}^{\rm Z-jet} ( p_{T\, Z},  p_{T\, \rm Jet}; R, \omega_{\rm min}   )   =  
\frac{
\frac{d\sigma_{AA}}{ d  p_{T\, Z} d p_{T\, \rm Jet} } }
{ \langle  N_{\rm bin}  \rangle
\frac{d\sigma_{pp}}{ d p_{T\, Z} d p_{T \, \rm Jet} }  }
  \;. \qquad
\label{RAAjet}
\end{eqnarray}

The nuclear modification factor provides a compact way through which to quantify 
the effects of the nuclear medium. Our predictions for 
$R_{AA}^{\rm Z-jet} ($R = 0.4$, \omega_{\rm min} = 20$~GeV)  
are presented  in Fig~\ref{raa_lhc_tagged}, where we plot the result that  
includes both collisional and radiative energy loss effects. Since part of the parton shower
energy is redistributed outside of the jet cone radius, the jets are pushed to lower values of 
$p_{T \rm Jet}$. This redistribution results in an enhancement in $R_{AA}^{\rm Z-jet}$ 
in the region of $ p_{T \, \rm Jet} < p_{T \, Z}$ and suppression in  $R_{AA}^{\rm Z-jet}$ 
in the region of $ p_{T \, \rm Jet} > p_{T \, Z}$, which is characteristic of in-medium  tagged-jet 
dynamics~\cite{Neufeld:2010fj}.

Next, we consider the $Z^0$-tagged jet event asymmetry, which is  obtained from Eq.~(\ref{ass})  
with  $p_{T\,1} \rightarrow p_{T\, Z}$, $p_{T\,2} \rightarrow p_{T\, \rm Jet}$.
Changing  variables from $(p_{T\, Z},p_{T\, \rm Jet})$ to $(A_J,p_{T\, \rm Jet})$ 
and then integrating over $p_{T\, \rm Jet}$,  we can express the differential 
$A_J$ distribution as follows:
\begin{eqnarray}
\frac{d\sigma}{dA_J}&=& 
\int_{p_{T\, \rm Jet\, min}}^{p_{T\,\rm Jet \,max}} dp_{T\, \rm Jet}    \frac{2 p_{T\,\rm Jet}}{(1-A_J)^2}
\frac{d\sigma}{ dp_{T\,Z}dp_{T\,\rm Jet}} \;, \quad
\label{ajcalc}
\end{eqnarray}
Here, $p_{T\, Z \, \rm min,max }$,   $p_{T\, \rm Jet \, min,max }$,   
can be specified by the experiment and determine the range of the $A_J$ distribution.
In the examples that follow  $p_{T\, Z  } \in (80,100)$~GeV and 
$p_{T\, \rm Jet  } > 20$~GeV.
In p+p collisions the asymmetry is sensitive to the ${\cal O}(G_F\alpha_s^n), \, n \geq 3$ 
multi-parton QCD processes. To ${\cal O}(G_F\alpha_s)$  $d\sigma/dA_J = \sigma \delta(A_J)$.    
In A+A collisions the modified $A_J$ distribution  provides information on the 
medium modification to parton shower evolution and jet propagation in the QGP.

\begin{figure}[!t]
\centerline{
\includegraphics[width = 0.95\linewidth]{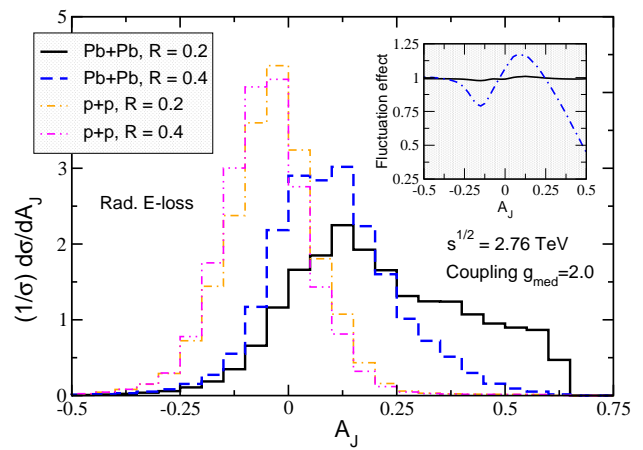}
}
\caption{The $Z^0/\gamma^*$-tagged jet event asymmetry for p+p collisions at  
$\sqrt{s} = 2.76$~TeV for two different $R=0.2, 0.4$. Predictions for central 
Pb+Pb collisions with radiative  medium-induced energy loss are also presented. 
Inset shows the effect of fluctuations in the background subtraction.
}
\label{radius}
\end{figure}

The tagged jet asymmetry distribution for p+p  collisions at 
$\sqrt{s} = 2.76$~TeV for two different $R=0.2, 0.4$ (dot-dashed and dot-dot-dashed curves) 
is shown in Fig.~\ref{radius}.
It peaks just below zero~\cite{Neufeld:2010fj}  and the mean values of $A_J$, defined as
\begin{equation}
\langle A_J \rangle = \int dA_J \, A_J \, \frac{1}{\sigma} \frac{d\sigma}{dA_J} \;, 
\label{mean}
\end{equation}
are shown in Table~\ref{stats}. 
The Pb+Pb curves (solid and dashed) are strongly shifted to $ A_J > 0 $ and considerably
broader than the p+p curves. This forward shift reflects the in-medium 
modification of the parton shower, which lowers the observed $p_{T \rm Jet}$ in 
Eq.~(\ref{ass}). For medium-induced parton splitting~\cite{Ovanesyan:2011kn},    
energy is lost due to large-angle radiation out of the jet reconstruction parameter 
$R$~\cite{jetty,jets}. The dependence upon $R$ shown in Fig.~\ref{radius}
further demonstrates this point, as the width and the average asymmetry of the 
curves with the smaller  radii are larger.

In a heavy-ion collision, jet reconstruction is complicated by an enormous 
soft hadronic background~\cite{Aad:2010bu}. When this background is subtracted
on average, its fluctuations can affect the $Z^0/\gamma^*$-tagged cross section. 
The result can be expressed as follows:
\begin{equation}
\frac{d\sigma_{AA}^{\rm fluc.}}{ d p_{T\, Z} d p_{T \, \rm Jet} } 
= \int d\delta p_T \,  \frac{d\sigma_{AA}(p_{T \, \rm Jet} - \delta p_T) }
{ d p_{T\, Z} d p_{T \, \rm Jet} } 
 \, \mathcal{N}(\delta p_T; \Delta p_T ^2) \;.
\label{smear}
\end{equation}
In Eq.~(\ref{smear})  $\mathcal{N}$ is a normal distribution. The ALICE experiment
has measured the standard deviation, which scales with the jet area, as $\Delta p_T \approx 11$~GeV for jet $R=0.4$ in
central Pb+Pb collisions at the LHC~\cite{Abelev:2012ej}. The inset of Fig.~\ref{radius} demonstrates the effect of these background fluctuations.
Specifically, the curves show the ratio of the result without background fluctuations 
to that with background fluctuations. Even for large radii, the effect is $<20\%$
in the region where  $A_J$ is significant. For small radii, such as $R=0.2$, 
the effect of fluctuations is completely negligible. This is demonstrated further 
quantitatively in Table~\ref{stats}, where we present the average asymmetry, 
$\langle A_J \rangle$, for a wide range of parameters both with and without the fluctuations 
(effect $<5\%$). 

\begin{figure}[!t]
\centerline{
\includegraphics[width = 0.95\linewidth]{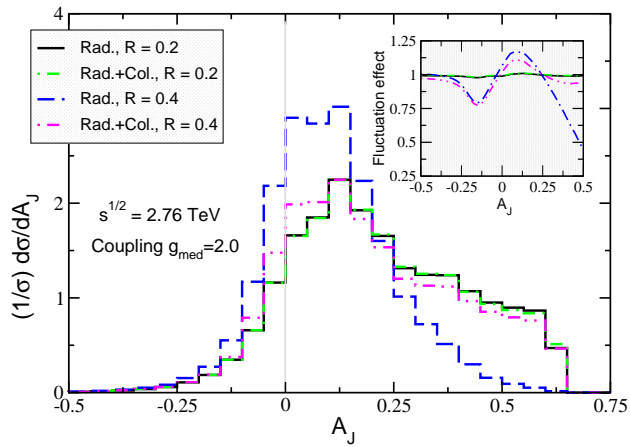}
}
\vspace*{-.2cm}
\caption{The asymmetry of $Z^0/\gamma^*$-tagged jet events ($R=0.2, \, 0.4$) 
for Pb+Pb collisions at $\sqrt{s}=2.76$~TeV with radiative and  radiative+collisional
 medium-induced energy losses. The collisional energy loss has a much more pronounced 
effect for larger radius. $A_J=0$ is shown to guide the eye. Inset shows the effect
of fluctuations in background  subtraction.  
}
\label{collision}
\end{figure}

We present the $Z^0/\gamma^*$-tagged jet event asymmetry for central  Pb+Pb collisions 
with radiative (solid and dashed curves)  and radiative+collisional  
(dot-dashed and dot-dot dashed curves)   medium-induced energy 
losses in Fig.~\ref{collision}. The collisional energy 
loss has a more pronounced effect in the curve with the larger radius. This occurs 
because collisional energy loss from a parton shower comes primarily from the 
radiated gluons, as demonstrated in \cite{Neufeld:2011yh}.  With the smaller 
radius most of the gluons are already outside of the jet cone making the extra energy 
loss redundant.  We  point out that background fluctuations again have minimal 
effect when the collisional energy loss is included, as can be checked 
from the insert in Fig.~\ref{collision} and more quantitatively, in Table~\ref{stats}.

\begin{table}[b!]
\vspace*{-.2cm}
\caption{Mean $A_J$ with and without  background fluctuations. 
Radius, jet-to-medium coupling and type of energy loss dependencies are
presented. } % title of Table
\centering % used for centering table
\begin{tabular}{c c c c c c  c} % centered columns (4 columns)
\hline\hline %inserts double horizontal lines
 System &   $\langle A_J \rangle_{\rm no\;fluct.}$ &   $\langle A_J \rangle_{\rm fluct.}$
    \\ [0.5ex] % inserts table
%heading
\hline   % inserts single horizontal line
p+p with R=0.2   &-0.025 & -0.025 \\[.5ex]
p+p with R=0.4   &-0.040 & -0.040 \\[.5ex] 
Pb+Pb, rad, R=0.2, $g_{\rm med}$=1.8  & 0.190 & 0.189 \\[.5ex] 
Pb+Pb, rad, R=0.2, $g_{\rm med}$=2.0   & 0.229 & 0.228 \\[.5ex] 
Pb+Pb, rad, R=0.2, $g_{\rm med}$=2.2   & 0.274 & 0.272 \\[.5ex] 
Pb+Pb, rad, R=0.4, $g_{\rm med}$=2.0   & 0.115 & 0.132 \\[.5ex] 
Pb+Pb, rad+col, R=0.2, $g_{\rm med}$=2.0   & 0.229 & 0.229 \\[.5ex] 
Pb+Pb, rad+col, R=0.4, $g_{\rm med}$=2.0   & 0.211 & 0.214 \\[.5ex] 
\hline %inserts single line
\end{tabular}
\label{stats} % is used to refer this table in the text
\end{table}

In summary, we presented the first study of the transverse momentum asymmetry of 
$Z^0/\gamma^*$-tagged 
jet events in $\sqrt{s}=2.76$~TeV reactions at the LHC. Our results are also 
qualitatively representative 
of other electroweak boson-tagged jet final states. We found both considerable broadening of the event 
asymmetry distribution and a characteristic shift of its peak to $A_J > 0$ in central Pb+Pb collisions  
relative to p+p collisions. Both features show very little sensitivity to the fluctuations of 
the underlying soft hadronic background and are related to the 2D nuclear modification factor 
$R_{AA}^{\rm Z-jet}$. Largely unaffected by cold nuclear matter effects, they can be used to accurately 
characterize the parton shower modification due to final-state interactions in the QGP.   

\small{{\it Acknowledgments}: This work was supported in part by the 
US Department of Energy, Office of Science, under Contract No. DE-AC52-06NA25396
and the LDRD program at LANL.}

\end{document}